\documentclass[final,1pt,times,twocolumn]{elsarticle}

\usepackage{amsmath}
\usepackage{graphicx}
\usepackage{tabularx}
\usepackage{multirow}
\usepackage{subcaption}
\usepackage{placeins}
\usepackage{caption}
\usepackage{adjustbox}
\usepackage{parskip}
\usepackage{hyperref}
\usepackage[dvipsnames]{xcolor}
\usepackage{caption}
\usepackage{subcaption}
\usepackage{ragged2e}
\usepackage{makecell}
\usepackage{titlesec}

\usepackage{amssymb}
\usepackage{amsmath}
\journal{Nuclear Physics B}

\usepackage{titlesec}
\titlespacing*{\section}{0em}{0.5em}{0.3em}
\titlespacing*{\subsection}{0em}{0.2em}{0.1em}
\titlespacing*{\subsubsection}{0em}{0.2em}{0.1em}
\makeatletter
\g@addto@macro\normalsize{%
  \setlength{\abovedisplayskip}{0.2em}
  \setlength{\belowdisplayskip}{0.2em}
  \setlength{\abovedisplayshortskip}{0.2em}
  \setlength{\belowdisplayshortskip}{0.2em}
}
\makeatother

\DeclareCaptionFormat{notsotiny}
{%
    \fontsize{6.00}{4}\selectfont #3
}

\DeclareCaptionFormat{tinytiny}
{%
    \fontsize{5.0}{4}\selectfont #3
}

\DeclareCaptionFormat{tiny}
{%
    \fontsize{6}{4}\selectfont #3
}

\DeclareCaptionFormat{medium}
{%
    \fontsize{6.5}{4}\selectfont #3
}

\DeclareCaptionFormat{cropfig}
{%
    \fontsize{5.75}{4}\selectfont #3
}

\DeclareCaptionFormat{cliccropfig}
{%
    \fontsize{5}{3}\selectfont #3
}

\newcommand{\preset}[1]{
    ``\textit{#1}''
}

\def\clicsamplewidth{\columnwidth}
\def\cliccolumnwidth{0.23\textwidth}

\def\cliccaptionspacing{-0.25em}

\newcommand\clicgtcropfig[3]{\begin{subfigure}[t]{1.0\textwidth}
                    \centering 
                    \setlength{\fboxsep}{0pt}
                    \fbox{\includegraphics[width=\textwidth]{#1}}
                    \captionsetup{aboveskip=0.2em, format=cliccropfig, justification=centering}
                    \caption*{Ground Truth}
                    \vspace{\cliccaptionspacing}
                    \caption*{Original size:}
                    \vspace{\cliccaptionspacing}
                    \caption*{#3}
                \end{subfigure}}

\newcommand\cliccropfig[6]{\begin{subfigure}[t]{1.0\textwidth}
                    \centering 
                    \setlength{\fboxsep}{0pt}
                    \fbox{\includegraphics[width=\textwidth]{#1}}
                    \captionsetup{aboveskip=0.2em, format=cliccropfig, justification=centering}
                    \caption*{#2}
                    \vspace{\cliccaptionspacing}
                    \caption*{#3bpp, #4db}
                    \vspace{\cliccaptionspacing}
                    \caption*{#5, #6}
                \end{subfigure}}

\begin{document}
\begin{frontmatter}



\title{PaaF: Raising the perceived quality of INR-Based Image Compression}


\author{Lorenzo Catania}
\affiliation{organization={Department of Mathematics and Computer Science, University of Catania},
            addressline={lorenzo.catania@unict.it}, 
            city={Catania},
            country={Italy}}
            
\author{Dario Allegra}
\affiliation{organization={Department of Mathematics and Computer Science, University of Catania},
            addressline={dario.allegra@unict.it}, 
            city={Catania},
            country={Italy}}
 
%% Abstract
\begin{abstract}
Implicit Neural Representations (INRs) have recently emerged as a promising paradigm for image compression, offering a fundamentally different approach from traditional and learned codecs. Nevertheless, INR-based methods for image compression suffer from long encoding times and a consistent performance gap in classic quality metrics such as PSNR. In this work, we explore the potential of purely INR-based compression methods and we propose PaaF (Picture as a Function), a novel INR-based image codec that introduces improved architectural design, adaptive quantization, and an efficient entropy coding scheme. These components are designed to enhance rate-distortion performance while preserving the simplicity and parallelizability of INR-based decoding.  Experimental results demonstrate consistent improvements over existing INR-based methods in both quantitative metrics and perceptual quality. These findings highlight the potential of INR-based approaches and contribute to narrowing the gap between functional representations and more established compression paradigms.

%Implicit Neural Representations (INRs) have recently proven to be effective in data compression tasks, offering an alternative to complex hand-crafted encoding and decoding pipelines. 
%However, INR-based methods for image compression suffer from long encoding times and a consistent performance gap in classic quality metrics such as PSNR. 
%In this work, we present PaaF, a novel image compression codec based on Implicit Neural Representations, which advances the current state-of-the-art NIF codec by consistently improving the rate-distortion ratio while maintaining comparable compression and decompression speed.
% We evaluate our proposal against INR-based, traditional and hybrid image codecs, considering quantitative metrics, computational times, and perceived visual fidelity.
% Our results show that PaaF enhances the INR-based compression performance by approximately 25\%  while reducing information loss.
% In addition, this novel codec excels in modern metrics such as LPIPS, outperforming or matching well-established and contemporary methods by offering superior visual fidelity even at very low bitrates. 
% An extensive ablation study highlights the contribution of each component of the proposed design, leading the way for further developments and a deep understanding of the properties and capabilities of INR-based methods.
\end{abstract}


% \begin{highlights}
% \item Deep overview into the strengths and weaknesses of Implicit Neural Representation for image compression
% \item Proposal of a set of improvements over existing INR-based codecs, including enhanced training methods and parameter compression techniques
% \item Presentation of PaaF, a novel INR-based codec based on the proposed architecture, released as open-source for public use
% \item PaaF outperforms previous works on the field and compares or improves the results of hybrid overfitted codecs
% \end{highlights}

%% Keywords

\end{frontmatter}



\section{Introduction}

Image compression constitutes a core challenge in multimedia systems and is conventionally addressed using compression algorithms that underpin codecs such as JPEG~\citep{wallace1992jpeg} and AVIF~\citep{barman2020evaluation}. 
However, this approach has drawbacks, including a lengthy development process and the need for efficient heuristics. An alternative strategy is to use deep learning techniques to train neural networks to exploit image redundancies~\citep{balle2016end}. This strategy also has limitations, such as impractical computational requirements~\citep{Pan2021}. A third approach involves Implicit Neural Representations (INRs)~\citep{mildenhall2020nerf, Sitzmann}, where data is interpreted as functions mapping coordinates to features, approximated by neural networks.

In imaging, a picture is represented as a function mapping pixel coordinates to colors, offering advantages such as decoupling image content from resolution and enabling arbitrary resolution rendering. INRs decoding consists of decompressing network parameters and inferring pixel values, a process easy to parallelize and implement efficiently. 
Initial works on INR-based image compression have shown potential by tying traditional codecs like JPEG and WebP but suffered from encoding times in the order of hours~\citep{Dupont2021, Strumpler}. 
Some alternatives adopted techniques such as meta-learning~\citep{Strumpler} and hybridization with learned latents~\citep{ladune2023coolchic}, that improved performance to some extent but introduced constriction to specific training data or complex serial entropy coding schemes that burden the decoding process.

% In this work, we propose a set of architectural choices for INR-based systems, an adaptive quantization and entropy coding scheme, and a variety of training procedures tuned for different applications. 
% These components are integrated in PaaF, the proposed INR-based image codec that advances the state-of-the-art and compares well with existing methods in terms of quantitative metrics, while exhibiting a consistent visual fidelity compared to the alternatives, especially at low bitrates.

In this work, we investigate the design space of Implicit Neural Representations (INRs) for image compression, focusing on advancing purely INR-based codecs and we propose PaaF (Picture as a Function). 
This novel codec departs from previously existing solutions, leveraging improved architectural choices, an adaptive quantization that completely replaces the traditional fixed-bits strategy and an economized entropy coding scheme. 
Our objective is to strengthen the performance of INR-based methods within this emerging paradigm, rather than comparing against fundamentally different compression frameworks.

%In this work, we propose PaaF, a purely INR-based image codec powered by innovative architecture choices, an adaptive quantization and entropy coding scheme. It is important to note that our paper specifically focuses on advancing purely INR-based codecs, aiming to improve their competitiveness within this emerging paradigm rather than comparing against fundamentally different compression frameworks.
%Our proposal advances the state-of-the-art in terms of quantitative metrics and improves visual fidelity compared to existing alternatives.
% The contributions presented in this paper are summarized in the following:
% \begin{itemize}
%     \item Discussion of the advantages and limitations of INR-based image codecs, and the proposal of an alternative network architecture and improved training methods and parameter compression techniques.
%     \item A novel image codec named PaaF based on INRs evaluated against existing INR-based codecs.
%     \item Availability of the codec as open-source and publicly accessible: \href{https://github.com/aegroto/nifv2}{https://github.com/aegroto/nifv2}.
% \end{itemize}
The contributions presented in this paper are summarized in the following:
\begin{itemize}
    \item Discussion of INR-based image codecs, and the proposal of innovatory technical components for such systems.
    \item A novel image codec named PaaF based on INRs evaluated against existing INR-based codecs.
    \item Availability of the codec as open-source. \textcolor{blue}{The full source code and results will be shared on GitHub upon publication.}
\end{itemize}

% The remainder of this paper is structured as follows.
% Section~\ref{sec:related_works} introduces the research subject with an overview of the state-of-the-art methods based on implicit representations and hybrid overfitting.
% % In Section~\ref{sec:background} we provide the basic concepts and the formalisms regarding implicit neural representations adopted in the this study.
% In Section~\ref{sec:method} we provide the basic concepts and the formalisms regarding implicit neural representations adopted in this study.
% Then, we present PaaF, the proposed implicit codec for image compression, designed on innovative insights and architectural choices.
% In Section~\ref{sec:experiments}, we evaluate the codec on several datasets against a range of baselines, validating the value of its components through an extensive ablation study.
% Alongside quantitative evaluations in terms of quality metrics and execution times, we analyse multiple image samples in-depth, assessing the perceived quality of the reconstruction, as well as the artifacts introduced by the different codec categories.
% Section~\ref{sec:conclusions} concludes the paper with a summary of the results and findings, discussing their impact on implicit image compression. Finally, we provide some hints for future research in the field.

The remainder of this paper is structured as follows. 
Section~\ref{sec:related_works} introduces the research subject with an overview of the state-of-the-art.
Section~\ref{sec:method} provides basic concepts and formalisms of INRs. We then present PaaF, our proposed codec for image compression. 
Section~\ref{sec:experiments} evaluates the codec on several datasets against various baselines, including quantitative and qualitative evaluations and an ablation study. 
Section~\ref{sec:conclusions} concludes the paper with a summary of results, discusses their impact, and provides hints for future research.

\section{Related work}\label{sec:related_works}
% The representation of data as neural networks has been formerly explored by seminal works on 3D shapes~\citep{park2019deepsdf, Mescheder_2019_CVPR} and radiance fields~\citep{mildenhall2020nerf}.
% The following works have focused on improving the capacity of such networks to represent high-frequency details. 
% This has been achieved by adding positional encodings on input features~\citep{Tancik2020} or using more expressive activation functions such as sinusoids~\citep{Sitzmann} and wavelets~\citep{saragadam2022wire}.
% These simple architectures were adopted in the first works about multimedia compression using INRs, such as COIN~\citep{Dupont2021} \textit{(COmpression using Implicit Neural representations)} for images and NeRV~\citep{Chen2021} \textit{(NEural Representation for Videos)}.
% These works have demonstrated that end-to-end overfitted neural networks compare with widely adopted codecs, avoiding complex handcrafted encoding pipelines.
% Although these works laid the foundations for further research on INR-based multimedia compression, they suffer from suboptimal compression ratios and prohibitively long processing times.

The representation of data via neural networks was first explored in 3D shapes~\citep{park2019deepsdf} and radiance fields~\citep{mildenhall2020nerf}. Building on these early efforts, later works improved high-frequency detail capture through input positional encodings~\citep{Tancik2020} or expressive activations like sinusoids~\citep{Sitzmann} and wavelets~\citep{saragadam2022wire}. Consequently, early INR-based compression methods, such as COIN~\citep{Dupont2021} for images and NeRV~\citep{Chen2021} for videos, demonstrated that end-to-end overfitted networks could rival traditional codecs by eliminating handcrafted pipelines. Nevertheless, these foundational approaches faced limitations in compression efficiency and computational scalability.

% Sudden works improved on these first proposals by meta-learning common parameters~\citep{Strumpler, coinpp} and expressing signals in terms of modulations.

% In the case of images, important contributions were given by Strumpler et al.~\citep{Strumpler}, which use better quantization schemes and positional encodings to improve the compression, and NIF~\citep{nif2023} \textit{(Neural Imaging Format)}, which adopts novel training strategy and network architecture reducing the encoding times while comparing or improving previous proposals in terms of rate-distortion ratio.
% A recent study line hybridizes functional data representations with learned features fed as input to small neural networks.
% The ground-breaking work Instant-NGP~\citep{mueller2022instant} performs fitting in terms of seconds on high-resolution data through mapping input coordinates to multi-resolution hash grids. However, the compression task of such features has been only partially tackled by following works~\citep{Girish_2023_ICCV, mahmoud2023cawanerf}.
% Regarding images, COOL-CHIC~\citep{ladune2023coolchic} and derived methods~\citep{kim2023c3, leguay2023lowcomplexity} are based on a similar architecture, using learned latents instead of positional features as input to a tiny neural network.
% These hybrid methods obtain state-of-the-art rate-distortion ratios at the cost of a pixel-wise serial decoding process, which increases the computational complexity, although they consistently reduce the number of MAC operations \textit{(Multiply-ACcumulate)} per decoded pixel.
For images, key advancements include Str\"umpler et al.~\citep{Strumpler}, which improved compression via better quantization and positional encodings, and NIF~\citep{nif2023}, which introduced novel training and architecture to reduce encoding times while maintaining competitive rate-distortion performance. 
Our proposed PaaF codec improves over existing works on a wide range of aspects, including quantitative perceptive metrics and reduced visual artifacts. We benchmark our implementation across diverse scenarios, demonstrating clear advantages over prior works.

\section{Proposed method}\label{sec:method}
In this section, we give an overview of the formalism and of the main concepts behind the design of PaaF.

\subsection{Coordinates-based image representation}
A key challenge in INRs is identifying a coordinate-to-sample mapping that facilitates the neural network fitting.
% In the case of images, this kind of continuous representation is opposed to the classic interpretation of a picture as a raster of pixels of width and height $(W, H)$.
In the case of images, given the position of a pixel $p$ in the raster as a couple $(p_x, p_y)$, where $p_x \in \{0, 1, ..., W-1\}$ and $p_y \in \{0, 1, ..., H-1\}$,  the resulting functional image representation $I$ is a mapping in the form:
\begin{align}
    (c_x, c_y) &= \left(\frac{2p_x}{W-1} - 1, \frac{2p_y}{H-1} - 1\right) \\
    I(c_x, c_y) &= (p_r, p_g, p_b) 
\end{align}
Where $(c_x, c_y)$ are referred to as the coordinates of pixel $p$ with RGB values $(p_r, p_g, p_b)$.

\subsection{Colorspace transform and channel-wise mean displacement}
% The target image is cast to the YCbCr colorspace before beginning any training phase.
% Then, the mean value of each channel is subtracted from the values of each pixel. After this transform, the target values $t$ of each pixel $p$ are:
The target image undergoes YCbCr conversion and per-channel mean subtraction, yielding transformed pixel values $t$ for each pixel $p$.
Given the original RGB values $p_r, p_b, p_g$ and their YCbCr counterparts $p_y, p_{cb}, p_{cr}$ with per-channel means $m_y, m_{cb}, m_{cr}$ the applied transform is:
\begin{equation}
    (t_y, t_{cb}, t_{cr}) = (p_y, p_{cb}, p_{cr}) - (m_y, m_{cb}, m_{cr})
\end{equation}
% Where $p_r, p_b$ and $p_g$ are the RGB values of the original pixel, $p_y$, $p_{cb}$ and $p_{cr}$ are their counterparts in the YCbCr space and $m_y, m_{cb}$ and $m_{cr}$ are the mean values for each channel.
% We refer to the subtraction of the mean of each channel as \textit{channel-wise mean displacement}.
% This color transform makes it possible to weigh the influence of luminance and chrominance on the loss values during training and \textit{mean displacement} has proven to increase the reconstruction quality in our experiments.
% This transformation has been empirically proven to increase the reconstruction quality in our experiments.
% As done in previous works~\citep{Strumpler, nif2023}, these values are then normalized to the range $[-1, 1]$.
This transformation empirically improves reconstruction quality in our experiments. Following prior work~\citep{Strumpler, nif2023}, values are then normalized to $[-1, 1]$.
% \newcommand{\noveltyheader}[0]{\textit{Novelties compared to NIF~\citep{nif2023}:} }

% \noveltyheader In the previous method no transformation was applied to the target signal, which was fitted in the RGB space and without any mean displacement.

\subsection{Exponential configurations}
% Previous research \citep{nif2023} has empirically proven that reducing the number of features of deeper layers in the network improves the bits-per-pixel ratio, adding bottlenecks proportionally to the depth of the network.
% PaaF adopts a heuristic to ease the configuration of sequential hyperparameters, such that the i-th element of the sequence $h_i$ is exponentially decreased: 
Prior work~\citep{nif2023} demonstrates that reducing deeper-layer feature counts improves bits-per-pixel ratio by introducing depth-proportional bottlenecks. 
PaaF adopts an exponential heuristic for hyperparameter configuration:
\begin{equation}
    h_i = m + (M - m) * p^{s}, \quad p = 1 - i/L
\end{equation}
% Where $M$ and $m$ are the first and the last values of the sequence, respectively, $L$ is the length of the sequence, and $s$ is a factor which regulates the exponential reduction.
where $M$ and $m$ are the first/last sequence values, $L$ its length, and $s$ controls exponential reduction.
% \subsubsection{Architecture bottlenecks}
% To configure the shape of the neural network, $M$ and $m$ are set to the number of features in the first and the last layer, respectively.
% The bottlenecks are added exponentially with the depth of the current layer based on the hyperparameter $e$.
% Note that if $e = 1$, the number of features is linearly decreased and if $M = m$ then the network is analogous to a traditional MLP with equal hidden features.
% \subsubsection{Weight restarts}
% Analogously, the number of weights restarted and the range of the perturbation values are exponentially decreased during training, with $M$ and $m$ corresponding to the initial and final values.
% Note that this formulation is analogous to the one proposed in~\citep{nif2023}.
The network shape is configured by setting $M$ and $m$ to the feature counts in the first/last layers. Bottlenecks are added exponentially with layer depth via the hyperparameter $e$. Similarly, weight restarts and perturbation ranges decay exponentially during training, with $M$ and $m$ as initial/final values.

% \noveltyheader In NIF, the number of features was explicitly specified for each layer, complicating the configuration tuning process. 
% In PaaF, finding the best configurations for each setting took less time by using simple hyperparameter tuning techniques (e.g. Bayesian optimization).

\subsection{Fine-grained quantization and configuration search}\label{sec:quant}
% Quantization is crucial when compressing floating-point values as it permits representing them as discrete integers and reducing their entropy.
% Conventionally, parameter values are quantized by mapping them to integer symbols in the range $[-2^{b-1}, 2^{b-1}-1]$, where $b$ is the number of bits used to represent the quantized values.
% Although this approach maximizes the representation precision for the given amount of bits, it forces the maximum symbol to be a power of 2.
% In practice, given quantization support of $b$ bits, choosing a maximum symbol $M$ such that $2^{b-2} \leq M \leq 2^{b-1}$ does not reduce the number of bits needed to represent each symbol of the alphabet, but reduces the alphabet size and therefore the overall entropy.
% Also, previous works~\citep{ladune2023coolchic} have empirically observed how in this kind of network the weights of each layer are laplacian-distributed after training; therefore, a non-uniform quantization scheme may reduce the quantization error.
% Based on these insights, we propose the following formulas to quantize and dequantize, respectively: 
Quantization reduces floating-point entropy by mapping values to discrete integers, typically in $[-2^{b-1}, 2^{b-1}-1]$ for $b$-bit precision. While this maximizes bit efficiency, it constrains the maximum symbol to powers of 2. Relaxing this to $2^{b-2} \leq M \leq 2^{b-1}$ preserves bit-width but reduces alphabet size and entropy. Observations~\citep{ladune2023coolchic} show post-training weights follow a Laplacian distribution, motivating non-uniform quantization to minimize error. Our proposed quantization/dequantization formulas are:
\begin{equation}
    v_{q} = \Bigl\lfloor \left(\frac{v}{B}\right) ^ {\frac{1}{s}} * M \Bigr\rceil 
    \qquad
    v_{d} = \left( \frac{v_{q}}{M} \right) ^ {s} * B
\end{equation}
% where $v$ is the value to be quantized, $v_q$ and $v_d$ are the quantized and dequantized values respectively, $M$ is the number of symbols chosen to represent the module of each value, $B$ is the bound of original values and is set to their maximum such that floating-point values are normalized in the range $[0, 1]$ before quantization and brought back to their original interval on dequantization. At the same time, $s$ is a scale factor that balances the quantization precision of low and high values and, assuming values are laplacian-distributed around 0 with standard deviation $\sigma$, it is empirically calculated as $s = 1 + \frac{\sigma}{\sqrt{2}}$.
where $v$ is the input value, $v_q$/$v_d$ its quantized/dequantized forms, $M=2^e$ the symbol count, $B$ the original-value bound (normalizing floats to $[0,1]$ pre-quantization). 
The scale factor $s=1+\sigma/\sqrt{2}$ (for Laplacian-distributed values, $\sigma$ is the standard deviation) balances precision across magnitudes. 
% To optimize the rate-distortion tradeoff, the best \textit{entropy} value $e$, with $M = 2^e$, is chosen through the following dichotomic search algorithm:
Optimal $e$ is selected via dichotomic search to minimize rate-distortion:
% \begin{itemize}
%     \item Evaluate the rate-distortion ratio of $N$ samples from the search range $[e_l, e_r]$;
%     \item Select the best and the second-best performing values $e_b$ and $e_s$;
%     \item If the width of the search range $|e_b - e_s|$ is less than a given threshold $p$, then stop the algorithm and choose $e_b$. Otherwise, perform the search again in the range $[e_b, e_s]$
% \end{itemize}
\begin{itemize}
    \item Evaluate rate-distortion for $N$ samples in range $[e_l, e_r]$;
    \item Select best ($e_b$) and second-best ($e_s$) entropy values;
    \item Terminate if $|e_b - e_s| < p$ (return $e_b$); else repeat search in $[e_b, e_s]$.
\end{itemize}
Note that $e$ is not necessarily an integer value; therefore, $M$ is not limited to a power of 2.
The following estimation function $g(e)$, named \textit{gain}, is used to estimate the score of each sample:
\begin{equation}\label{eq:gain}
    g(e) = r_e * t - q_e 
\end{equation}
% Where $r_e$ is the rate reduction, $q_e$ is the quality drop, and $t$ is a hyper-parameter named $tolerance$ that defines the allowed quality drop in proportion to the rate reduction.
% $r_e$ and $q_e$ are calculated relative to the reference rate and distortion values, calculated in practice by quantizing the values with the maximum acceptable number of bits.
% The reason to search for $e$ instead of the actual value $M$ is that the rate grows logarithmically with $M$ and linearly with $e$, therefore, estimating the rate-distortion ratio is straightforward when using $e$ in Equation~\ref{eq:gain}.
Here, $r_e$ and $q_e$ denote rate reduction and quality drop (relative to a reference $b$-bit quantization), with hyperparameter $t$ ($tolerance$) balancing their tradeoff. Searching over $e$ (where $M=2^e$) simplifies optimization, as rate scales logarithmically with $M$ (linearly with $e$), enabling direct estimation in Equation~\ref{eq:gain}.

% \noveltyheader  NIF uses naive fixed 8-bit quantization. The proposed method, instead, achieves consistently better rate-distortion performance by finding for each image encoding the stricter quantization parameters that do not affect the reconstruction quality.

\subsection{Entropy coding and packing}
% The compressed representation of an image is interpreted as a stream of bytes with a structure known to the decoder.
% Metadata describing the image resolution is added at the beginning of the stream.
% To maximize the compression ratio, the quantized values of each parameter are entropy coded using \textit{Brotli}~\citep{alakuijala2018brotli} and the range coder from the \textit{constriction} library~\citep{bamler2022constriction} by estimating both a Laplacian and a Gaussian distribution. 
% The smallest of these entropy-coded streams is chosen and added to the stream along with a header describing the quantization and entropy-coding configuration.
The compressed image is byte-streamed with resolution metadata prefixed. Quantized parameters undergo entropy coding via Brotli~\citep{alakuijala2018brotli} and constriction’s range coder~\citep{bamler2022constriction}, evaluating both Laplacian and Gaussian distribution. The smaller coded stream is selected, with headers specifying quantization/entropy settings.

% \noveltyheader NIF limited entropy coding to straightforward brotli. While simple, this strategy does not always bring the smallest bitstream.

% \section{Proposed method}\label{sec:method}

\subsection{Encoding pipeline}
\begin{figure*}
    \centering
    \includegraphics[width=\textwidth]{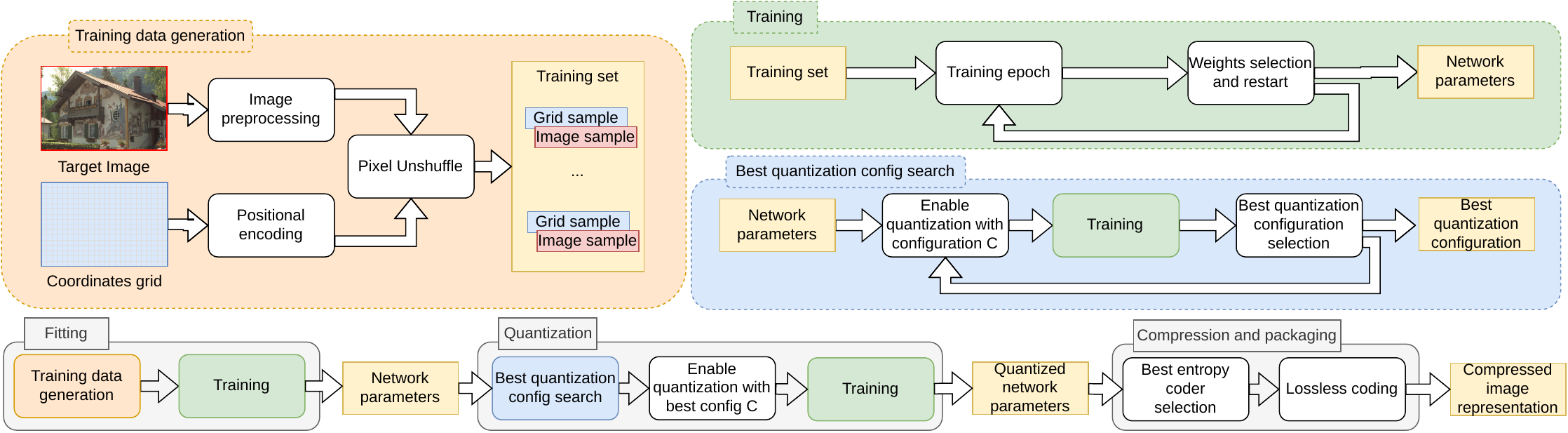}
    \caption{Overall scheme of the compression pipeline adopted in PaaF. Note that the weight restart is not performed during the last epoch of each fitting phase.}
    \label{fig:pipeline}
\end{figure*}

% First attempts at INR-based image compression \citep{Strumpler, Dupont2021} relied on training the neural network on the functional image mapping for a high number of iterations.
% This trivial method is penalized by requiring high computational power and memory, which grow proportionally to the image resolution and make it impossible to apply these solutions to high-resolution images. 
Early INR compression methods~\citep{Strumpler, Dupont2021} optimized functional image mappings via prolonged training, but scaled poorly with resolution-computational/memory demands rendered them impractical for high-resolution inputs.
% NIF~\citep{nif2023} proposed a better-structured training pipeline to apply heuristics such as weight perturbation between training epochs and to consistently decrease the computational requirements by reducing the portion of the image for each optimization step by pixel unshuffling \citep{Shi_2016_CVPR}.
% However, this method lacks proper formalization and does not take into account pre-processing and post-processing of the signal.
% NIF~\citep{nif2023} improved scalability via structured training (weight perturbation, pixel unshuffling~\citep{Shi_2016_CVPR}) to reduce per-step compute, but remained heuristic omitting the proper formalization of each component.
The first leap forward a structured training process was proposed by NIF~\citep{nif2023}, which pipeline introduced weight perturbation, and dataset subdivision by pixel unshuffling~\citep{Shi_2016_CVPR}, reducing per-step compute.
However, these heuristics were never properly formalized and generalized to ease exploration and configurations.
% In this paper, we describe an advancement to this pipeline, which solves some of the issues of existing alternatives and is easily generalizable to different data modalities, and a lean network architecture composed of a single module, unlike the original dual-sided NIF architecture.
This work advances prior pipelines, addressing key limitations while generalizing across modalities. Our architecture replaces NIF’s dual-sided design with a single lean module.
Figure \ref{fig:pipeline} gives an overview of the compression pipeline designed for our method.
% Before training the network, preprocessing steps such as color transform and channel-wise mean offsets are applied to the target image data and positional encoding is applied to the coordinates tensor.
% Analogously to NIF, the training set is built by applying pixel unshuffle with a factor $S$ to the processed image and coordinates tensor, considering each of the resulting $S^2$ tensors as a separate sample.
% The resulting training batch is used to perform a sequence of fitting epochs. At the end of each epoch, the parameters with the best loss value are selected, perturbed and fed to the next epoch.
% In contrast to the original NIF pipeline, quantization is applied when selecting the best-performing parameters.
% Then, the best quantization configuration is selected, performing fast-fitting epochs for each possible configuration following the strategy presented in section~\ref{sec:quant}. In the end, a full quantization-aware fitting phase is performed to obtain the final quantized parameters.
% These last are entropy-coded and packed into a compressed bitstream.
Pre-training, inputs undergo color transform, mean offsets, and positional encoding. Pixel unshuffling (factor $S$) splits data into $S^2$ samples. 
% Training proceeds via iterative fitting epochs: best-loss parameters are perturbed and carried forward, with quantization integrated into selection (unlike NIF). 
Training proceeds via iterative fitting epochs: best-loss parameters are perturbed and carried forward.
Quantization is then integrated into the selection at the end of each epoch. 
Optimal quantization config is chosen via fast-fitting and quantization selection as presented in ~\ref{sec:quant}, followed by a full quantization-aware fine-tuning. Final parameters are entropy-coded into the bitstream.

% \noveltyheader The presented pipeline is a generalization of the one presented in our previous work, plus includes some additional steps such as adaptive quantization.

\subsection{Network architecture}

\begin{figure}[t]
    \centering
    
    \begin{subfigure}[b]{0.8\columnwidth}
        \centering
        \includegraphics[angle=90, width=\textwidth]{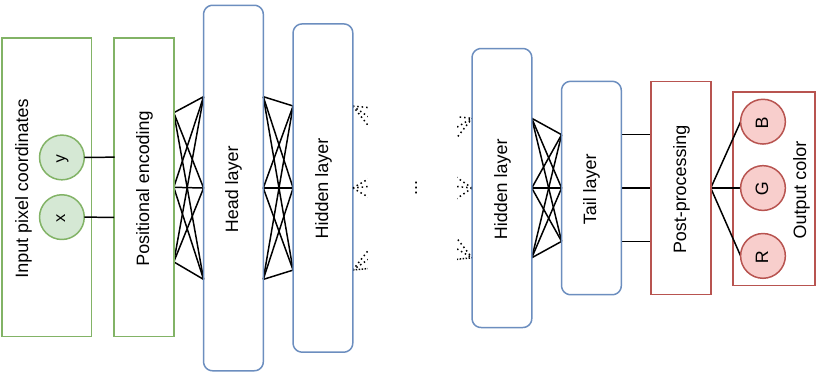}
    \end{subfigure}
    \begin{subfigure}[b]{0.8\columnwidth}
        \centering
        \includegraphics[angle=0, width=\textwidth]{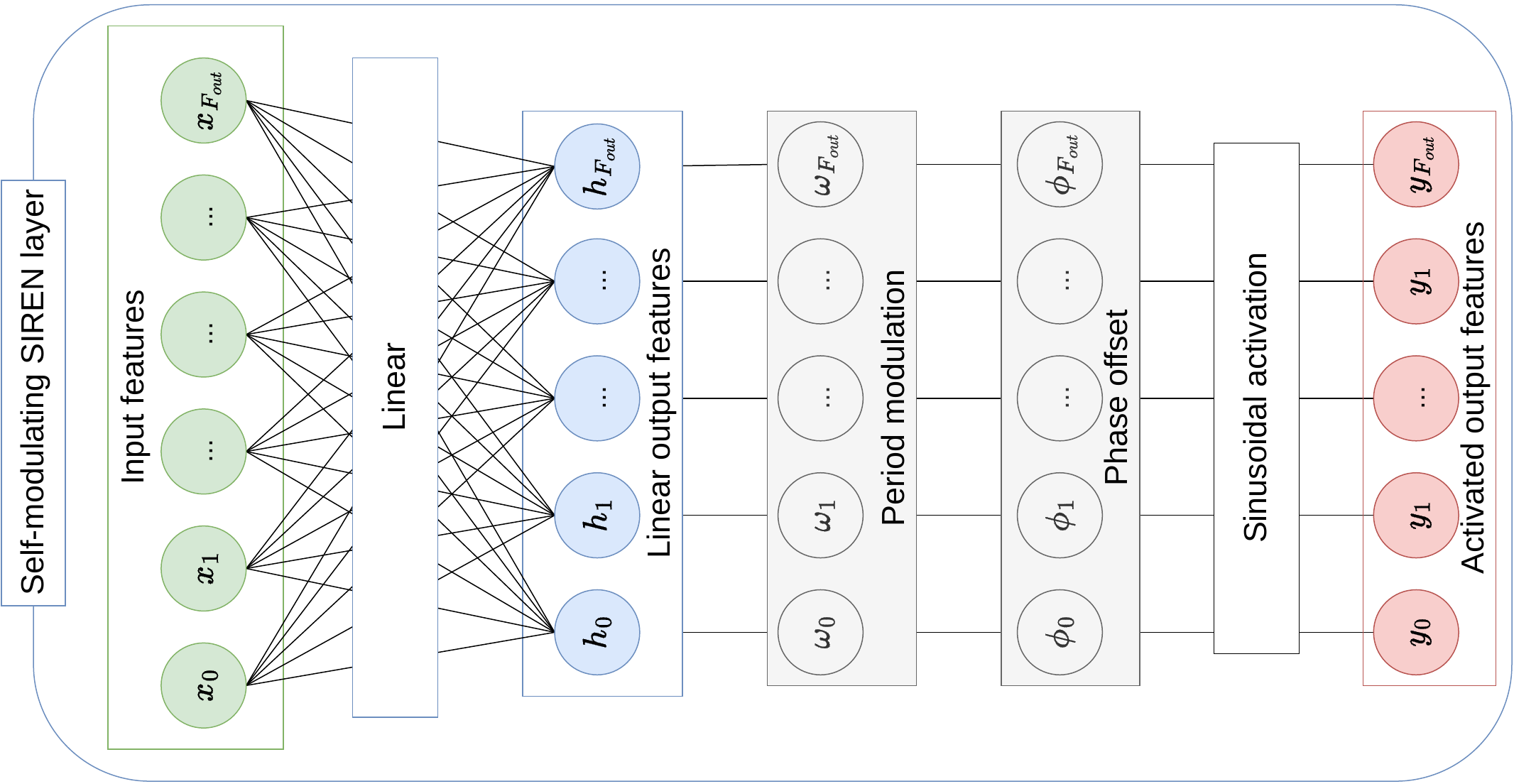}
    \end{subfigure}
    
    \caption{Architecture of the self-modulated layer (top) and the overall neural network (bottom) adopted in PaaF.}
    \label{fig:network_architecture}
\end{figure}

An overall view of the network architecture is given in Figure \ref{fig:network_architecture}. 
% Input coordinates are passed through a positional encoder, then passed to a multi-layer perceptron.
% Tail features are then passed to post-processing filters, which revert pre-processing steps and give the actual RGB pixel values as output.
% Note that, in line with the concept behind INRs, the decoder shares the same architecture as the encoder.
% After training is done on the encoder's end, the parameters are serialized and transmitted to the decoder, which copies them into its own instance of the model and infers to decode the frame.
% The advantage of this architecture compared to the original NIF~\citep{nif2023} one is that only a single forward pass is required for each pixel, and no separate modulation network is used. %, increasing both encoding and decoding speed.
% Instead, each layer is a self-modulating SIREN~\citep{Sitzmann} layer \textit{(SInusoidal REpresentational Network)}, except for the last layer, named tail, where no activation is performed.
Note that, in line with the concept behind INRs, the decoder shares the same architecture as the encoder.
Coordinates undergo positional encoding before processing by a multi-layer perceptron. 
The last layer is named tail, and its features pass through post-processing filters to reconstruct RGB values.
After training, parameters are serialized for decoder replication. 
This architecture enables single-pass pixel inference, eliminating NIF’s~\citep{nif2023} modulation network. All layers (except the tail) are followed by self-modulating SIREN~\citep{Sitzmann} activations.

\subsubsection{Self-modulating sinusoidal activations}
% In their classical formulations, SIRENs~\citep{Sitzmann} are traditional multi-layer perceptrons with sinusoidal activations, where the weights are initialized following a precise scheme. 
% The periods of sinusoidal activations are increased by multiplying the output features of each layer by a hyper-parameter $\omega_0$:
Traditional SIRENs~\citep{Sitzmann} employ sinusoidal MLPs with structured weight initialization, scaling layer outputs by $\omega_0$ to control activation periods:
\begin{align}
    y_l &= sin(\omega_0 * (W_l * y_{l-1}^T + B_l))
    \addtocounter{equation}{1}\tag{\theequation}
\end{align}
Where $W_i, B_i$ are the weights and the biases of the l-th layer, and $y_l$ is the output of the l-th layer.
% Period modulations have proven to be an effective technique to improve the representation capabilities of SIRENs \citep{mehta2021modulated, nif2023}, but existing techniques make use of additional modules to be added in the network architecture, increasing the overall computational overhead.
While period modulation enhances SIREN representational power~\citep{mehta2021modulated, nif2023}, prior methods introduce auxiliary modules, increasing computational cost.
% In this formulation, $\omega_0$ and $\omega_0 * B_i$ represent the periods and the phases of the sine activation respectively.
% We propose an alternative sinusoidal activation layer where the phase of each feature is deterministically sampled to guarantee enough variability around the interval $[-\pi, \pi]$, removing the bias on each linear layer.
% Analogously, periods are modulated such that each output feature is activated by using a slightly different period than $\omega_0$, removing the separate modulation network adopted until now to introduce variation over periods of each activation.
% As the domain of the phase is limited, we replace the traditional learned biases with fixed values sampled following a sinusoidal distribution which we refer to as phases. 
% Also, we associate a period to every output feature of each hidden layer such that 
Here, $\omega_0$ and $\omega_0*B_i$ define activation periods/phases.
Our proposed self-modulating layer deterministically samples phases in $[-\pi, \pi]$ to eliminate linear bias and modulates periods per-feature around $\omega_0$, obviating auxiliary modulation networks.
Therefore, the i-th phase $\phi_i$ and period $\omega_i$ of the activations of a layer having $F_{out}$ output features are calculated as:
\begin{equation}
    \phi_i = sin(2\pi\Psi_i) * \pi, \quad \omega_i = \omega_0 * \Omega_i
\end{equation}
where $\psi_i$ is the i-th point of a linear space with interval $[-\Psi, \Psi]$ and size equal to $F_{out}$, $\Omega_i$ is sampled analogously and $\Psi$ and $\Omega$ are hyperparameters that define the ranges from where values are sampled.
% Note that these parameters are not learned and can be sample during inference on the decoder by using the same pseudo-randomizer of the encoder; therefore, there is no need to transmit them.
% By fixing these values, each output feature in a layer is bound to a specific phase, and the network is limited to learning the weights that vary the period of each input feature. 
These non-learned parameters are regenerable via sharing encoder/decoder pseudo-RNG settings, eliminating the need to transmit them. 
Fixed phases constrain per-feature period learning to weight adaptation only.
As a result, the calculations at each activation are given by the formula:
\begin{equation}
    y_i = sin(\omega_i * h_{i} + \phi_i)
\end{equation}
Where $h_{i}$ is the $i-th$ output feature of the linear transformation of the given layer.
The structure of each self-modulating layer is represented in Figure \ref{fig:network_architecture}.
In summary, the hyper-parameters to be set are the sampling ranges $\Psi$ and $\Omega$ and the base period $\omega_0$.
Weights are initialized likewise to traditional SIRENs based on the hyper-parameter $\omega_0$.

% \noveltyheader NIF dual architectures required two forward passes for each sample, while this novel architecture enables inference with a single pass. 
% Plus, this does not imply sacrificing reconstruction accuracy nor needing more trainable parameters.

\subsubsection{Loss functions}\label{sec:loss_functions}
% Analogously to other works in the field of learned image compression~\citep{xie2021}, we propose different loss functions that optimize the method for a specific objective. 
% In all formulas, $y$ is the original image and $\hat{y}$ is the reconstructed image.
Following learned compression approaches~\citep{xie2021}, we tailor loss functions to specific objectives, with $y$ and $\hat{y}$ denoting original/reconstructed images:
For PSNR, the loss function is given by a mathematically stable approximation of $log(cosh(x))$: 
\begin{equation}
    \small
    L_{PSNR}(y, \hat{y}) = \frac{\sum{((y - \hat{y}) + sp(2 *  (y - \hat{y})) - ln(2))}}{N}
\end{equation}
Where $sp$ is the SoftPlus function and is calculated as:
\begin{equation}
    sp(x) = ln(1+e^{x})
\end{equation}

% When looking for optimal MS-SSIM~\citep{ms-ssim} \textit{(Multi Scale Structural SIMilarity)}, we use a combination of L1 and SSIM~\citep{ssim} \textit{(Structural SIMilarity)}:
When looking for optimal MS-SSIM~\citep{ms-ssim}, we use a combination of L1 and SSIM~\citep{ssim}:
\vspace{0.2em}
\begin{equation}
    L_{MSSSIM}(y, \hat{y}) = \alpha_{L} * L1(y, \hat{y}) + \alpha_{S} * DSSIM(y, \hat{y})
\end{equation}

We also propose a \preset{Visual} preset adopting the same loss of NIF~\citep{nif2023}, which uses a combination of LogCosh as in $L_{PSNR}$ and SSIM, and has been shown to obtain good visual results:
\vspace{0.2em}
\begin{equation}
    L_{Vis}(y, \hat{y}) = \alpha_{L} * LogCosh(y, \hat{y}) + \alpha_{S} * DSSIM(y, \hat{y})
\end{equation}

$\alpha_{L}$ and $\alpha_{s}$ are values which balance the weight of each factor in the total loss and $DSSIM(y, \hat{y}) = 1 - SSIM(y, \hat{y})$. 

% \noveltyheader Our previous proposal was strictly tuned on a preset similar to the \preset{Visual} proposed above. PaaF, instead, exhibits improved versatility with its variegated available presets.

% \begin{figure*}
%     \centering
%     \includegraphics[width=\textwidth]{pipeline.pdf}
%     \caption{Overall scheme of the compression pipeline adopted in PaaF. Note that the weight restart is not performed during the last epoch of each fitting phase.}
%     \label{fig:pipeline}
% \end{figure*}

\section{Experiments}\label{sec:experiments}
We have evaluated our proposal on the Kodak~\citep{kodak} and the CLIC2020~\citep{CLIC2020} validation datasets, in line with previous works~\citep{ladune2023coolchic}.
We compare our results with the other purely functional INR-based methods COIN~\citep{Dupont2021}, ICE~\citep{ice}, Strumpler et al.~\citep{Strumpler} and NIF~\citep{nif2023}. We purposely restrict our comparison to purely INR-based methods to ensure a fair evaluation within the same modeling paradigm, as hybrid and latent-based learned codecs rely on fundamentally different encoding mechanisms.
Note that we also include JPEG as a reference traditional codec, not as a direct competitor, to provide an intuitive baseline for interpreting distortion characteristics at comparable bitrates.
%We also add JPEG for comparison with a traditional approach.
% We also add traditional codecs AVIF, JPEG, JPEG XL~\citep{jpegxl} and the autoencoder-based standard JPEG~AI~\citep{jpeg-ai} for comparisons with other different approaches.
% Note that reported COOL-CHIC results are obtained through an image-wise greedy configuration search, encoding every image with different time-consuming configurations and choosing the best result.
% As stated by the authors~\footnote{\url{https://github.com/Orange-OpenSource/Cool-Chic/issues/5}}, this process may take hours for CLIC2020 images with a gain of approximately 2\% in BD-rate.
%In contrast, PaaF results are obtained with single-shot encoding using the same configuration at a given bitrate for the whole dataset.

% We adopt the classic PSNR and the more advanced MS-SSIM~\citep{ms-ssim} and LPIPS~\citep{lpips} \textit{(Learned Perceptual Image Patch Similarity)} as evaluation metrics. 
% In line with previous works \citep{nif2023}, MS-SSIM values are normalized as $-10log_{10}(1-value)$ to scale them logarithmically, easing the comparisons of lines in plots.
% The adaptive quantization algorithm optimizes PSNR for the \preset{PSNR} preset and MS-SSIM for \preset{MS-SSIM} and \preset{Visual} presets.
% A comprehensive resume of the functions used in each preset is given in Table~\ref{tab:presets}.

We adopt PSNR, MS-SSIM~\citep{ms-ssim}, and LPIPS~\citep{lpips} as evaluation metrics. Following previous works~\citep{nif2023}, MS-SSIM values are normalized as $-10\log_{10}(1-\text{value})$ for logarithmic scaling, facilitating plot comparisons. Adaptive quantization optimizes PSNR for the \preset{PSNR} preset and MS-SSIM for the \preset{MS-SSIM} and \preset{Visual} presets.

% \begin{table}[]
%     \centering
%     \begin{tabular}{|c|c|c|}
%          \hline
%          Preset & Loss function & Quantization metric  \\
%          \hline
%          \textit{PSNR} & $L_{PSNR}$ & PSNR \\
%          \hline
%          \textit{MS-SSIM} & $L_{MSSSIM}$ & MS-SSIM \\
%          \hline
%          \textit{Visual} & $L_{Vis}$ & MS-SSIM \\
%          \hline
%     \end{tabular}
%     \caption{Resume of functions each in the configuration of each preset. Loss functions correspond to the ones described in Section~\ref{sec:loss_functions}, while quantization metric is used to evaluate the best entropy value in the adaptive algorithm described in Section~\ref{sec:quant}.}
%     \label{tab:presets}
% \end{table}

\subsection{Quantitative evaluation}
\begin{figure}[t]
    \centering
    \includegraphics[width=0.9\columnwidth]{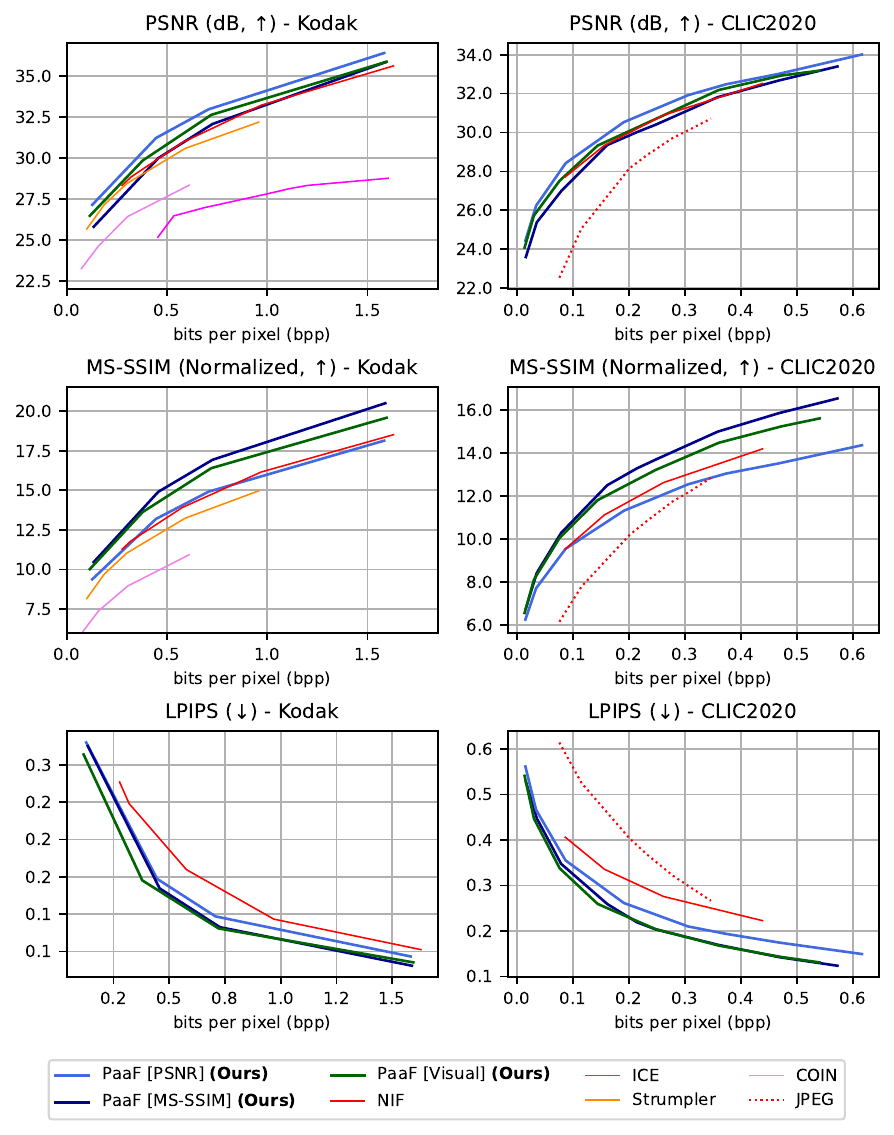}
    \caption{Quantitative experiments on the Kodak and CLIC2020 datasets. The preset used for PaaF is specified around square brackets.}
    \label{fig:quantitative-inr}
\end{figure}

\begin{table}[b]
    \centering
    \resizebox{\columnwidth}{!}{%
    \begin{tabular}{|c||c|c|c|c|c|c|}
        \hline
        \multirow{2}{*}{Anchor} &
        \multicolumn{3}{c|}{\textbf{Kodak}} &
        \multicolumn{3}{c|}{\textbf{CLIC2020}} \\
        \cline{2-7}
        & PSNR & MS-SSIM & LPIPS & PSNR & MS-SSIM & LPIPS \\
        \hline
        COIN     & \textcolor{OliveGreen}{-68.52\%} & \textcolor{OliveGreen}{-74.54\%} & - & \textcolor{OliveGreen}{-68.36\%} & \textcolor{OliveGreen}{-75.55\%} & \textcolor{OliveGreen}{-87.07\%} \\
        \hline
        Strumpler & \textcolor{OliveGreen}{-36.34\%} & \textcolor{OliveGreen}{-48.91\%} & - & \textcolor{OliveGreen}{-31.12\%} & \textcolor{OliveGreen}{-47.43\%} & \textcolor{OliveGreen}{-70.06\%} \\
        \hline
        NIF      & \textcolor{OliveGreen}{-25.72\%} & \textcolor{OliveGreen}{-37.48\%} & \textcolor{OliveGreen}{-39.07\%} & \textcolor{OliveGreen}{-19.19\%} & \textcolor{OliveGreen}{-33.74\%} & \textcolor{OliveGreen}{-51.39\%} \\
        \hline
    \end{tabular}
    }
    \caption{BD-Rate gains of the proposed PaaF against baseline methods on the Kodak and CLIC2020 datasets. Metrics for which the baseline results were not available are reported with "-".}
    \label{tab:quantitative_bdrate}
\end{table}

Results on the Kodak and CLIC2020 datasets are plotted in Figure~\ref{fig:quantitative-inr}.
The proposed PaaF model is compared to baseline methods across both the Kodak and CLIC2020 datasets. 
Our proposed PaaF outperforms other INR-based methods, across all metrics and bitrates, on the Kodak dataset.
On the CLIC2020 dataset, PaaF’s advantage is especially pronounced in LPIPS.
While PaaF surpasses COIN, Strumpler, and ICE, NIF remains competitive only under the PSNR preset for MS-SSIM.
This stems from NIF’s loss function, identical to PaaF’s Visual preset, ensuring balanced metric performance.

BD-Rate comparisons are shown in Table~\ref{tab:quantitative_bdrate}, with missing values for baseline methods indicated by -.
Following prior BD-Rate studies~\citep{bd_bible}, both MS-SSIM and LPIPS are log-scaled before BD-Rate calculations for fair comparison.
On the Kodak dataset, PaaF achieves substantial reductions, outperforming COIN and Strumpler by a significant margin. 
PaaF considerably improves over NIF, the best-performing baseline,  with a BD-PSNR of -25.72\% and a remarkable BD-LPIPS of -39.07\%.
The advantages of PaaF are further underscored in the CLIC2020 dataset, where it consistently surpasses all baselines across all metrics. 
In this wide high-resolution dataset, PaaF keeps distances with its competitors, reaching BD-MS-SSIM and BD-LPIPS rates from -50\% to -87\%.

\subsection{Qualitative comparisons}
\begin{figure*}
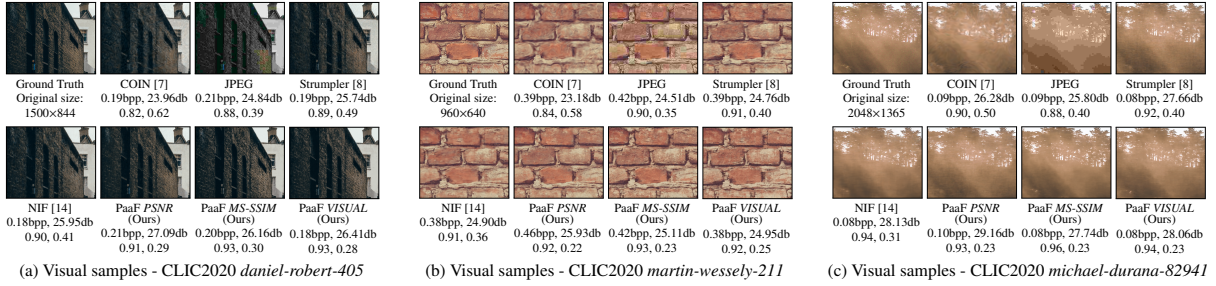

    \centering
    \resizebox{\textwidth}{!}{%
        \begin{tabular}{ccc}
            \begin{minipage}[t]{0.33\linewidth}
                \vspace{0pt}
                \input{visual_clic2020_daniel-robert-405_1.00_source}
            \end{minipage}
            &
            \begin{minipage}[t]{0.33\linewidth}
                \vspace{0pt}
                \input{visual_clic2020_martin-wessely-211_1.00_source}
            \end{minipage}
            &
            \begin{minipage}[t]{0.33\linewidth}
                \vspace{0pt}
                \input{visual_clic2020_michael-durana-82941_1.00_source}
            \end{minipage}
        \end{tabular}
    }
    \caption{Visual comparisons on CLIC2020 image details. Reported metrics are, in order, bits-per-pixel, PSNR, MS-SSIM and LPIPS.}
    \label{fig:visual_clic2020}
\end{figure*}

We selected sample images from the CLIC2020 dataset where visual artifacts and quality differences may not be captured by quantitative metrics.
In Figure~\ref{fig:visual_daniel-robert-405}, JPEG's aggressive quantization causes severe color aliasing in daniel-robert-405, while COIN and Strumpler fail to represent high-frequency details like wall edges. PaaF \preset{Visual} achieves a faithful scene reconstruction, reducing the blurriness compared to NIF.
A similar effect is evident in the brick details of martin-wessely-211 (Figure~\ref{fig:visual_martin-wessely-211}). PaaF representations appear superior, avoiding the blur and wavy artifacts present in other methods.
The \textit{Visual} preset's effectiveness is clear in the challenging details of michael-durana-82941 (Figure~\ref{fig:visual_michael-durana-82941}), such as trees and light rays, compressed at a low bitrate (0.08–0.10 bpp). COIN, Strumpler, and the \textit{PSNR} preset introduce significant noise, whereas the \textit{Visual} preset faithfully preserves both tree leaves and light shapes.

\subsection{Ablation study}
% \subsubsection{Quantization scheme}
% \begin{figure}
%     \begin{subfigure}[b]{0.45\textwidth}
%         \includegraphics[width=\textwidth]{ablations_quantization_psnr.pdf}
%         \includegraphics[width=\textwidth]{ablations_quantization_msssim.pdf}
%         \includegraphics[width=\textwidth]{ablations_quantization_visual.pdf}
%         \caption{Quantization scheme}
%         \label{fig:ablation-quantization}
%     \end{subfigure}
    
%     \begin{subfigure}[b]{0.45\textwidth}
%         \includegraphics[width=\textwidth]{ablations_loss_psnr.pdf}
%         \includegraphics[width=\textwidth]{ablations_loss_msssim.pdf}
%         \caption{Loss function}
%         \label{fig:ablation-loss}
%     \end{subfigure}
%     \caption{PaaF Ablation study on the Kodak dataset about the quantization scheme~\ref{fig:ablation-quantization} and loss function~\ref{fig:ablation-loss}}
% \end{figure}

\begin{table}[]
    \centering
    \resizebox{1.0\columnwidth}{!}{%
    \begin{tabular}{|c|c|c|c|}
\hline
\multicolumn{4}{|c|}{BD-Rate vs. [\%]}\\
\hline
Anchor                   & PSNR                & MS-SSIM             & LPIPS               \\
\hline
\multicolumn{4}{|c|}{Adaptive quantization scheme ablation}\\
\hline
Fixed 6-bits quantization&\textcolor{OliveGreen}{-8.48\%}&\textcolor{OliveGreen}{-12.67\%}&\textcolor{OliveGreen}{-3.00\%}\\
\hline
Fixed 7-bits quantization&\textcolor{OliveGreen}{-1.51\%}&\textcolor{OliveGreen}{-1.19\%}&\textcolor{OliveGreen}{-3.43\%}\\
\hline
Fixed 8-bits quantization&\textcolor{OliveGreen}{-8.13\%}&\textcolor{OliveGreen}{-5.98\%}&\textcolor{OliveGreen}{-11.24\%}\\
\hline
Fixed 12-bits quantization&\textcolor{OliveGreen}{-39.88\%}&\textcolor{OliveGreen}{-38.28\%}&\textcolor{OliveGreen}{-42.27\%}\\
\hline
\multicolumn{4}{|c|}{Loss function ablation}\\
\hline
PaaF [PSNR] w. L2 loss  &\textcolor{OliveGreen}{-2.37\%}&\textcolor{OliveGreen}{-36.47\%}&\textcolor{OliveGreen}{-15.27\%}\\
\hline
PaaF [PSNR] w. L2 + SSIM loss&\textcolor{OliveGreen}{-55.39\%}&\textcolor{OliveGreen}{-24.28\%}&\textcolor{OliveGreen}{-20.80\%}\\
\hline
PaaF [MS-SSIM] w. L2 loss&\textcolor{OliveGreen}{-5.05\%}&\textcolor{OliveGreen}{-37.88\%}&\textcolor{OliveGreen}{-21.66\%}\\
\hline
PaaF [MS-SSIM] w. L2 + SSIM loss&\textcolor{OliveGreen}{-57.82\%}&\textcolor{OliveGreen}{-25.86\%}&\textcolor{OliveGreen}{-26.95\%}\\
\hline
    \end{tabular}
    }
    \caption{BD-rates comparisons regarding the ablation study on the Kodak dataset}
    \label{tab:ablation_bdrates}
\end{table}

% Results for replacing the proposed quantization scheme search algorithm with a classical fixed scheme on the Kodak dataset are reported in Figure~\ref{fig:ablation-quantization}.
% BD-rates results in Table~\ref{tab:ablation_bdrates} confirm that the adaptive quantization scheme improves the rate-distortion ratio overall.
% Quantizing with 6 or 12 bits is underperforming in every case. 
% Both 7 and 8-bit quantization alternatively obtain results comparable to our proposed scheme at low bitrates but are overcome when encoding with higher quality, demonstrating the effectiveness of our proposed algorithm, which adapts along bitrate changes.
% The gap is most evident when optimizing for MS-SSIM on the \textit{MS-SSIM} preset and in the \textit{PSNR} preset for every metric.
BD-rates results in Table~\ref{tab:ablation_bdrates} confirm that the adaptive quantization scheme enhances the overall rate-distortion ratio. Quantization with 6 or 12 bits consistently underperforms. While 7 and 8-bit quantization yield results comparable to the proposed scheme at low bitrates, they are outperformed at higher quality encoding, highlighting the adaptability of our algorithm across varying bitrates. The performance gap is most pronounced when optimizing for MS-SSIM on the \textit{MS-SSIM} preset and for every metric on the \textit{PSNR} preset.
% \subsubsection{Loss function}
% Various alternative loss functions have been evaluated for both PSNR and MS-SSIM optimized versions, with the results reported in Figure~\ref{fig:ablation-loss}. 
% BD-Rates reported in Table~\ref{tab:ablation_bdrates} make evident that our combination of loss functions surpasses the other combinations based on L2 and SSIM, especially in terms of LPIPS.
% The classic L2 (\textit{Mean Squared Error}) loss function obtains similar PSNR results compared to our proposed LogCosh function, but it is outperformed in any other term.
% A naive addition of an SSIM factor to L2 improves the result in terms of MS-SSIM, but causes a PSNR drop, and it still does not outperform our proposed L1+SSIM combination.
% Note that we do not consider the \preset{Visual} tuning configuration in this ablation as it differs from the \preset{MS-SSIM} configuration only by the loss itself.
% LogCosh+SSIM, which is the loss function proposed in NIF~\citep{nif2023}, it is outperformed by our proposals in terms of PSNR and MS-SSIM at all bitrates. In terms of LPIPS, it obtains slightly better results at higher compression ratios than our proposed MS-SSIM optimized version, but its efficiency consistently drops at higher bitrates. 
The same results show that our loss function combination outperforms L2 and SSIM-based ones, particularly in LPIPS. The classic L2 (MSE) loss matches our LogCosh function in PSNR but underperforms in all other metrics. Adding SSIM to L2 improves MS-SSIM but reduces PSNR and still falls short of our L1+SSIM combination. The \preset{Visual} tuning configuration is excluded from this ablation, as it only differs from \preset{MS-SSIM} by the loss function.

\section{Conclusion}\label{sec:conclusions}

This paper analyzes pure INR-based image compression. While learned methods can still outperform INR codecs, we significantly narrow the perceptual quality gap. We propose PaaF, an open-sourced implicit codec whose novel architectural and training designs advance purely functional INR compression both quantitatively and qualitatively. PaaF’s simple design enables easy implementation, fast decoding, and high-fidelity, artifact-free reconstructions even at low bitrates. Its main limitation is the need of per-image training; however, decoding is efficient and parallelizable via parameter decompression and coordinate inference. Future work entails computational optimizations and extensive subjective studies to validate perceived quality, building on Section~\ref{sec:experiments}. Ultimately, these results and growing community interest~\citep{Dupont2021, Strumpler, nif2023, ladune2023coolchic, Chen2021} highlight the potential of functional data representations and implicit compression to revolutionize multimedia.

\FloatBarrier

\bibliographystyle{elsarticle-num}
\bibliography{egbib}

\end{document}